\documentclass[12pt,dvips]{article}
\usepackage{rotating}
\usepackage{epsfig}
\usepackage{graphicx}
\usepackage{color}
\usepackage{cite}

\def\be{\begin {equation}}
\def\ee{\end {equation}}
\def\beqa{\begin {eqnarray}}
\def\eqa{\end {eqnarray}}
\def\ba*{\begin {eqnarray*}}
\def\ea*{\end {eqnarray*}}
\def\m{\mathrm}

\begin{document}

\def\thefootnote{\fnsymbol{footnote}}

\begin{flushright}
{\tt INFN/TC\_11/6}
{~}\\
\vspace{0.5cm}
January 2012
\end{flushright}

\begin{center}
{\bf {\Large The Spin Contribution to the Synchrotron Light}}
\end{center}

\medskip

\begin{center}{\large
{\bf Mario~Conte}
}
\end{center}

\begin{center}
{\em Dipartimento di Fisica dell'Universit\`a di Genova and
  INFN-Sezione di Genova, 
  Via Dodecaneso 33, 16146 Genova, Italy.}\\[0.2cm]

\end{center}

\bigskip

\centerline{\bf ABSTRACT}

\noindent  
In order to detect the spin contribution to the synchrotron radiation, the so-called spin light,
we propose to compare the characteristics of the radiation emitted by a spin-less charged particle
with the huge crop of data regarding the synchrotron light, i.e. the radiation emitted by a particle
endowed with a magnetic moment. Helium nuclei are proposed as the lightest stable spin-less
charged particles available.

\medskip
\noindent

\newpage

\section {Introduction}
The contribution \cite{BT,KBBG} of the particle magnetic moment to the synchrotron radiation has been mainly considered from the theoretic point of view. 
As experimental test we propose to compare the spectral characteristics of the radiations emitted by spin-less charged particles with the well known data regarding 
the synchrotron radiation. A subtraction between these two sets of results should evidence the role of the magnetic moment. The expression of the energy emitted 
over a revolution by a circulating particle of charge $q$ is
\begin{equation}
\delta U = \frac {q^2}{3\varepsilon_0}\frac {\gamma^4}{\rho} 
\label{1} 
\end{equation}
where $\varepsilon_0$ is the vacuum permittivity, $\gamma$ is the particle Lorentz factor and $\rho$ is the bending radius. The quantity (\ref{1}) can be obtained 
via arguments of classical electromagnetism and/or special relativity applied to an accelerated charged particle, without any mention to the role of an eventual spin. 
The well known bremsstrahlung is a typical phenomenon of this kind. Nevertheless, in order to be able to speak of {\bf synchrotron radiation} particles must \cite{SK} 
be ultra-relativistic (UR), move on a circular trajectory and irradiate an energy much smaller than their own total energy. Besides, the critical frequency 
\begin{equation}
\omega_c = \frac {3}{2} \frac {c}{\rho} \gamma^3 ,     
\label{2} 
\end{equation}
over which the spectrum of this radiation is centered, can be evaluated considering that this radiation is emitted within a narrow cone of aperture $1/\gamma$ 
and making use of the Nyquist theorem: i.e. via still classical arguments.  Therefore, we may conclude that eq. (\ref{1}) and eq. (\ref{2}), fundamental in the synchrotron 
radiation physics, can be also used for  spin-less charged particles.
 
\section {A Few Practical Examples}
The lightest spin-less charged particle which can be easily accelerated is the Helium-4 nucleus whose mass is

\begin{equation} 
m_{He_4} = 3.752 ~ {\m {GeV/c}}^2 = 6.691 \times 10^{-27} ~ {\m {kg}}           
\label{3} 
\end{equation}
A glance at eq. (\ref{2}) points out the difficulty of making the Helium ions emit a radiation inside regions where it is possible to perform accurate measures. In fact we 
should need extremely high energy machines in order to obtain a $\gamma$ sufficiently big. To this purpose, we work out a few realistic numerical values relative to 
RHIC \cite{RHIC} and LHC \cite{LHC}, chosen as the most energetic accelerator available. Recalling that UR particles fulfill the relation 
$\gamma \simeq \beta\gamma = \frac {p}{mc} = \frac {qB\rho}{mc}$, the Lorentz factor of interest is 
\begin{equation} 
\gamma = \frac {2eB\rho}{m_{He_4}c} ~~~ [e = {\m {elementary~charge}}] 
\label{4} 
\end{equation}
thinking of employing double ionized Helium. Then, picking up data from the references \cite{RHIC} and \cite{LHC} mentioned before, we can write the quantities 
shown in Table I.
\begin{table}
\centering
\caption{  Table I.
}
\vskip 0.1 in
\begin{tabular}{|c|c|c|c|c|c|c|c|} \hline
  & B [T] & $\rho$ [m] & $\gamma$ & $\frac {c}{\rho}$ [s$^{-1}$] &  $\omega_c$ [s$^{-1}$] &              $\hbar\omega_c$ & $\lambda$ \\
\hline
\hline      
{\bf RHIC} & 3.458 & 242.78 & 134 & 1.23$\times10^6$ & 4.46$\times10^{12}$ & 2.94 meV  &  0.42 mm \\
\hline
 {\bf LHC}  &   8.4  &  2780  & 3731 & 1.08$\times10^5$ & 8.40$\times10^{15}$ &  5.53 eV  &  224 nm \\
\hline
\end{tabular}
\label{datab}
\end{table}

\end {document}